\documentclass[a4paper,twoside]{article}

\usepackage{epsfig}
\usepackage{subcaption}
\usepackage{calc}
\usepackage{amssymb}
\usepackage{amstext}
\usepackage{amsmath}
\usepackage{amsthm}
\usepackage{multicol}
\usepackage{pslatex}
\usepackage{apalike}
\usepackage{SCITEPRESS}     

\begin{document}

\title{Noise-Resilient Automatic Interpretation of Holter ECG Recordings}

\author{\authorname{Egorov Konstantin\sup{1}, Sokolova Elena\sup{1}, Avetisian Manvel\sup{1} and Tuzhilin Alexander\sup{2}}
\affiliation{\sup{1}Sberbank AI Lab, Moscow, Russia}
\affiliation{\sup{2}New York University, New York City, USA}
\email{\{egorov.k.ser, sokolova.el.vladimirov, avetisian.m.s\}@sberbank.ru, atuzhilin@stern.nyu.edu}
}
 

\abstract{Holter monitoring, a long-term ECG recording (24-hours and more), contains a large amount of valuable diagnostic information about the patient. Its interpretation becomes a difficult and time-consuming task for the doctor who analyzes them because every heartbeat needs to be classified, thus requiring highly accurate methods for automatic interpretation. In this paper, we present a three-stage process for analysing Holter recordings with robustness to noisy signal. First stage is a segmentation neural network (NN) with encoder-decoder architecture which detects positions of heartbeats. Second stage is a classification NN which will classify heartbeats as wide or narrow. Third stage in gradient boosting decision trees (GBDT) on top of NN features that incorporates patient-wise features and further increases performance of our approach. As a part of this work we acquired 5095 Holter recordings of patients annotated by an experienced cardiologist. A committee of three cardiologists served as a ground truth annotators for the 291 examples in the test set. We show that the proposed method outperforms the selected baselines, including two commercial-grade software packages and some methods previously published in the literature.}

\onecolumn \maketitle \normalsize \setcounter{footnote}{0} \vfill

\section{\uppercase{Introduction}}

Cardiovascular diseases remain the leading cause of death throughout the world, according to the World Health Organization (WHO) \cite{WHO18}. Timely screening and diagnosis of these diseases can significantly reduce mortality caused by them.
An electrocardiogram (ECG) is one of the most affordable and common tools for recording heart rhythm that facilitates the diagnosis of wide range of heart pathologies. Given the fact that heart rhythm abnormalities and other malfunctions may occur irregularly, monitoring for an extended period of time is necessary to detect these events. Long-term ECG recordings (24-hours and more) contain a large amount of valuable diagnostic information about the patient, however their interpretation becomes a difficult and time-consuming task for the cardiologist who analyzes them. Holter ECG recordings contain hundreds of thousands of heartbeats and, ideally, position of each of them should be determined accurately and each heartbeat should be classified individually. 

The task of interpretation of long-term Holter recordings is challenging \cite{Schlapfer17}. In one study, computerized interpretation of ECG signals identified non-sinus rhythms with accuracy of only 53.5\% \cite{Shah07}. Another study \cite{Lindow19} found that among ECGs with a computer-based diagnosis of atrial fibrillation or atrial flutter, the diagnosis was incorrect in almost 10\%. In almost half of the cases, the misdiagnosis was not corrected by the over-reading physician. The clinical impact of the computer-based ECG misinterpretation was also evaluated in \cite{Bond18} where it was demonstrated that incorrect automated diagnosis (AD) significantly affects the reader's interpretation accuracy. In particular, diagnosis accuracies achieved by cardiology fellows dropped by 43.20\% when an incorrect AD was presented to them.

Duration, amplitude and morphology of QRS complex (RR interval, width of QRS complex and slopes of various segments) are important criteria for detection of abnormal heartbeats \cite{Osowski01}. Therefore, the purpose of our work is to create an automatic artificial intelligence algorithm for making the key ECG-measurements more precise on the long-term noisy recordings, such as detection of the heart beat positions (R-peak) and morphology of QRS complex (wide or narrow).
In this paper, we make the following contributions. First, we introduce a signal-wise Convolutional Neural Network (CNN) architecture that operates on the channel level. Second, we propose a novel method of how to add patient-specific information to our model by stacking the ECG segmentation and the patient-wise classification models (NN+GBDT). 

Third, we show that the proposed method outperforms the selected baselines, including two commercial-grade software packages and some methods previously published in the literature. Finally, we show that our method approaches the performance level of experienced cardiologists, which we demonstrate in an experiment involving 291 annotated ECG recordings and 3 highly skilled cardiologists.

\section{\uppercase{Related work}}

Historically, the automated ECG signal interpretation is implemented by expertly-created feature extraction algorithms (onset and offset of the different waves, measurements of various intervals, amplitude parameters etc.), while the classification is performed by decision rules, which are also created and tuned by experts. In order to improve the accuracy of such methods, a number of machine learning algorithms were applied to the problem. These methods allowed use of more informative features, e.g. time to frequency conversion methods (e.g. wavelet-transform) in order to extract features from variable-length waveforms \cite{Essam17}. However, the problem of automated interpretation of ECG signals still remains under-explored, despite all these efforts \cite{Kaplan18}\cite{Estes13}. Furthermore, with the advent of deep-learning based methods, new expectations have been developed recently that cardiologist-level interpretation of ECGs can be achieved by using modern deep neural networks \cite{Hong19}.

One such paper \cite{Shashikumar18} describes how a multi-stage model has been applied to the atrial fibrillation detection problem. In particular, the signal was split into 10-minute segments, and the noise reduction and frequency analysis using wavelet transform was performed on these signals. Furthermore, features were extracted from the spectrograms using a CNN, while a BRNN and attention layers managed to capture temporal patterns in the extracted features, resulting in the final classification layer calculating probabilities of predicting atrial fibrillation.

Another work in applying deep learning to the ECG interpretation task is presented in \cite{Rajpurkar17}, where a 34-layer CNN was able to exceed performance of a cardiologist in detecting a wide range of heart arrhythmias by leveraging a large annotated dataset and a very deep CNN. Our work differs from \cite{Rajpurkar17} as follows. First, the training and the test sets in the paper were constructed in a way to make it more balanced, i.e. taking only approximately 2 30-seconds windows for each patient, thus possibly making false positive metric inaccurate. Second, the paper works only for single-lead ECG records, while our work focuses on extracting features from \textit{multi-channel} ECG recordings.

\begin{figure*}[!h]
  \centering
   {\epsfig{file = 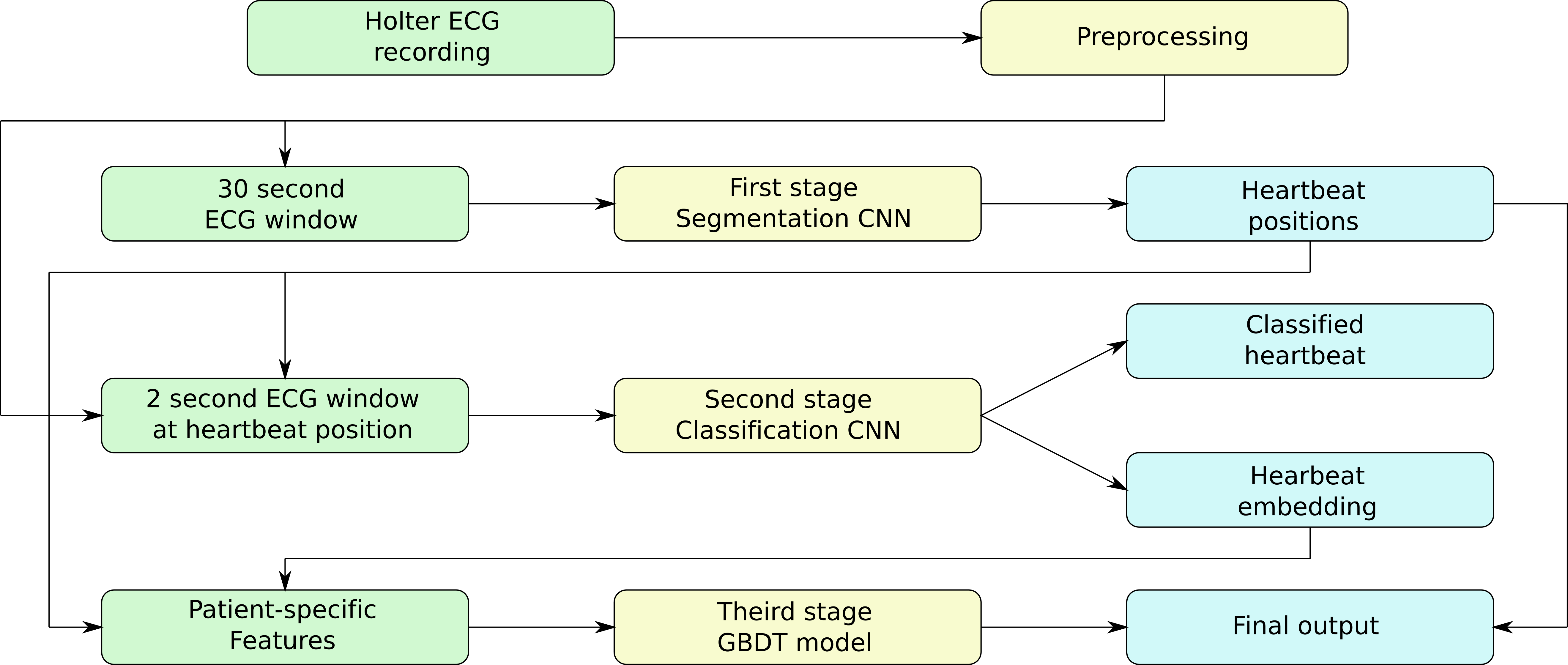, width = 15cm}}
  \caption{The whole algorithm of our approach. }
  \label{fig1}
\end{figure*}

The problem of noise in the ECG recordings is studied in \cite{Everss-Villalba17}. First, the authors defined 5 classes of noise, ordered by the clinical impact of the noise on the parameters to be measured in the ECG. Second, various measures of noise level have been proposed, such as baseline wander, powerline interference, and standard deviation noise (which was proposed in the paper). Third, a noise map is built, which characterizes temporal distribution of the noise. The paper demonstrates that noise level can be prohibitively high for automated or human interpretation of ECG signals.

Based on this literature review, we conclude that significant progress has been made in addressing the problem of automated ECG analysis over the last several years. However, the problem is far from being solved due to the complexities associated with having highly unbalanced datasets and prevalence of significant levels of noise in the ECG data.

\section{\uppercase{Data}}
\subsection{Dataset}

Data used in the present study was collected from different clinics of one of East European countries. We worked with a large dataset of ECG recordings consisting of 5,095 ambulatory 2--lead ECG recordings of 24--hour duration (Table 1 presents the statistics pertaining to this dataset). The ECG data is sampled at a frequency of 250 Hz. All the personally identifiable information has been deleted by the clinics. Each recording was annotated by experienced cardiologists using commercial ECG analysis software. The annotation contained positions of individual QRS-compleses, as well as class labels, such as wide or narrow QRS-complexes, arrythmia events (extrasistoles and pauses) and different types of pauses. Furthermore, these ECG recordings were contaminated by noisy and unreadable segments or one of the electrodes was disconnected for some time and then reconnected again. Thus, our task is building a model which is resilient to such levels of noise.

We split the dataset into the training, validation and test subsets in the proportion of 74\%, 20\% and 6\% respectively (having 3804, 1000 and 291 records in each subset). 
The test set was additionally annotated by a committee of three independent certified and practicing cardiologists, and the "ground truth" was determined by the voting of those experts.

\begin{table}[h]
\vspace{-0.2cm}
\caption{Dataset Statistics. }\label{tab1} \centering
\begin{tabular}{|c|c|c}
  \hline
  Number of records & 5 095 \\
  \hline
  Total length & 5232 days 21 hours 6 minutes \\
  \hline
  Total wide QRS count & 6 515 633 (1.2 \%) \\
  \hline
\end{tabular}
\end{table}

\subsection{Preprocessing}

In order to reduce noise in the signal, we applied the following preprocessing steps to our data. 
First, we fill signal from disconnected electrodes with linear interpolation between the last non-zero sample and the first non-zero sample after disconnection.
The goal of this step is to remove extreme high-frequency spikes in the points where the electrodes are disconnected and reconnected. 

Second, we removed the wandering trend and made isoline close to zero by subtracting two passes of the mean filter with a 100-samples window. Third,  we applied a low-pass filter with the cutoff frequency of 40Hz. The last prepossessing step is down-sampling with the factor of 2. 
During the training step, we also normalized the amplitude and subtracted the channel-wise median from each training window to ensure that isoline is as close to zero as possible.

\section{Method}
\subsection{Overview}

\begin{figure*}[t]
  \centering
   {\epsfig{file = 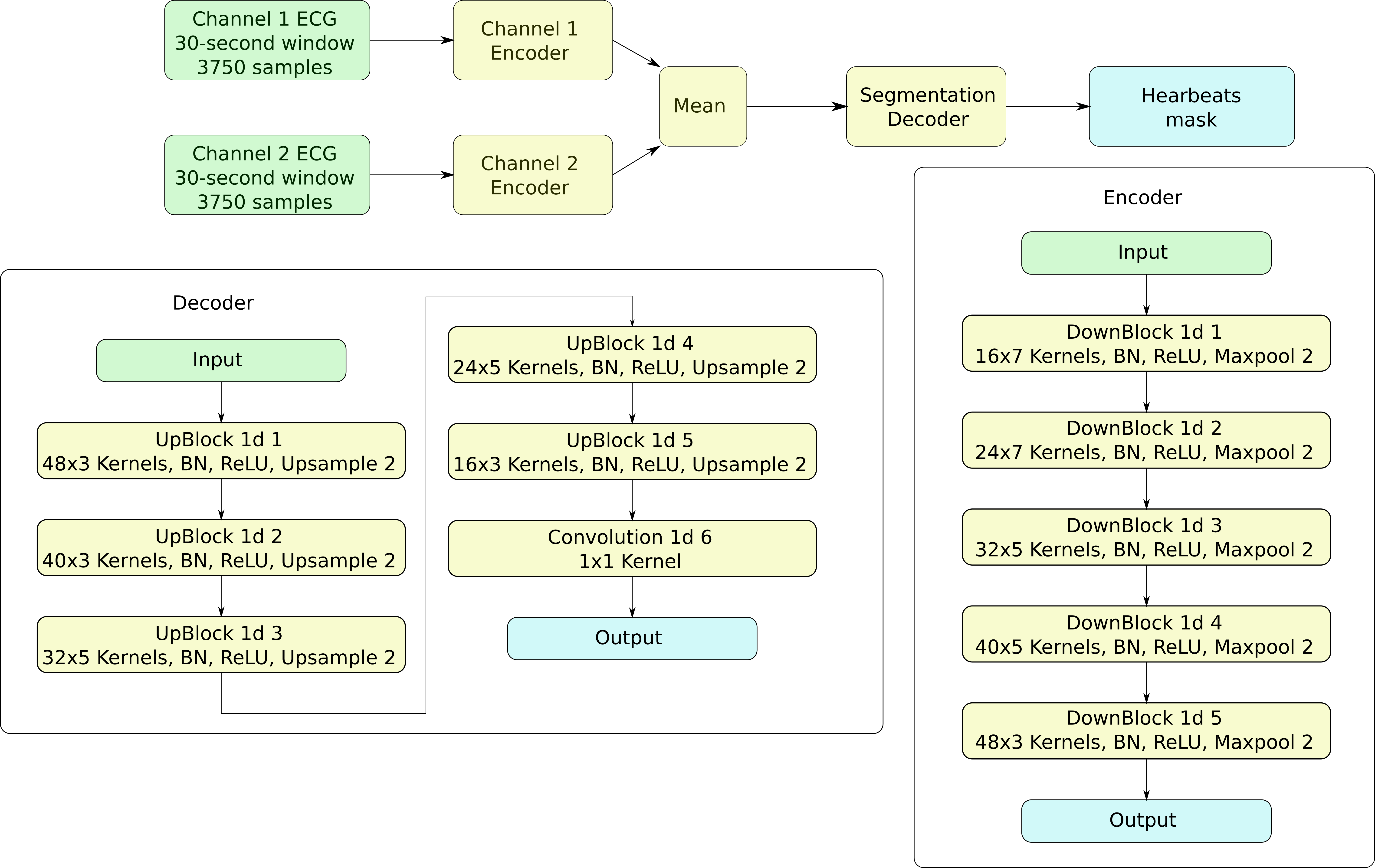, width = 15cm}}
  \caption{Neural network of the first stage. }
  \label{fig2}
\end{figure*}

Our approach consists of three main stages of analysis (see Fig.~\ref{fig1}). First stage is a CNN segmentation model with the encoder-decoder architecture with a bottleneck layer. There are two separate CNN encoders for each channel with identical structure, but different weights. Before bottleneck resulting feature maps of the encoders are averaged to ensure that both encoders generate similar features from different channels and are supplementary to each other (see Fig.~\ref{fig2}). This decision was based on the fact, that major parts of some recordings have periods (sometimes up to 100\% of time) when one of the electrodes is either detached or has a very low signal quality index (SQI) \cite{Li17}. While in  \cite{Shashikumar18} authors facing similar challenge are feeding the neural network only through the channels with highest SQI, we found that a more beneficial approach is to use all channels at all the times, but to let the neural network learn to distinguish the noisy or absent signals on its own. Our goal was to create a model that is working on any number of electrodes, and each attached electrode enhances the overall performance, while detachment of the electrode does not result in performance deterioration.

The second stage comprises a CNN with the structure similar to the first stage, but with the classification head instead of segmentation. Its input consist of the 2-second windows of the signal centered by position detected in the first stage.

Third stage comprises a gradient boosting decision tree classifier. We use it to enhance the performance of the heartbeat classification step by incorporating global patient features into the model. While the first-stage CNN processes only 30 seconds of the signal and the second-stage CNN processing 2 seconds of signal, we found that it is important to capture the individual characteristics of a given recording and a patient and use them in final classification. It turns out that this step significantly improves the overall classification process.

\begin{figure*}[h]
  \centering
   {\epsfig{file = 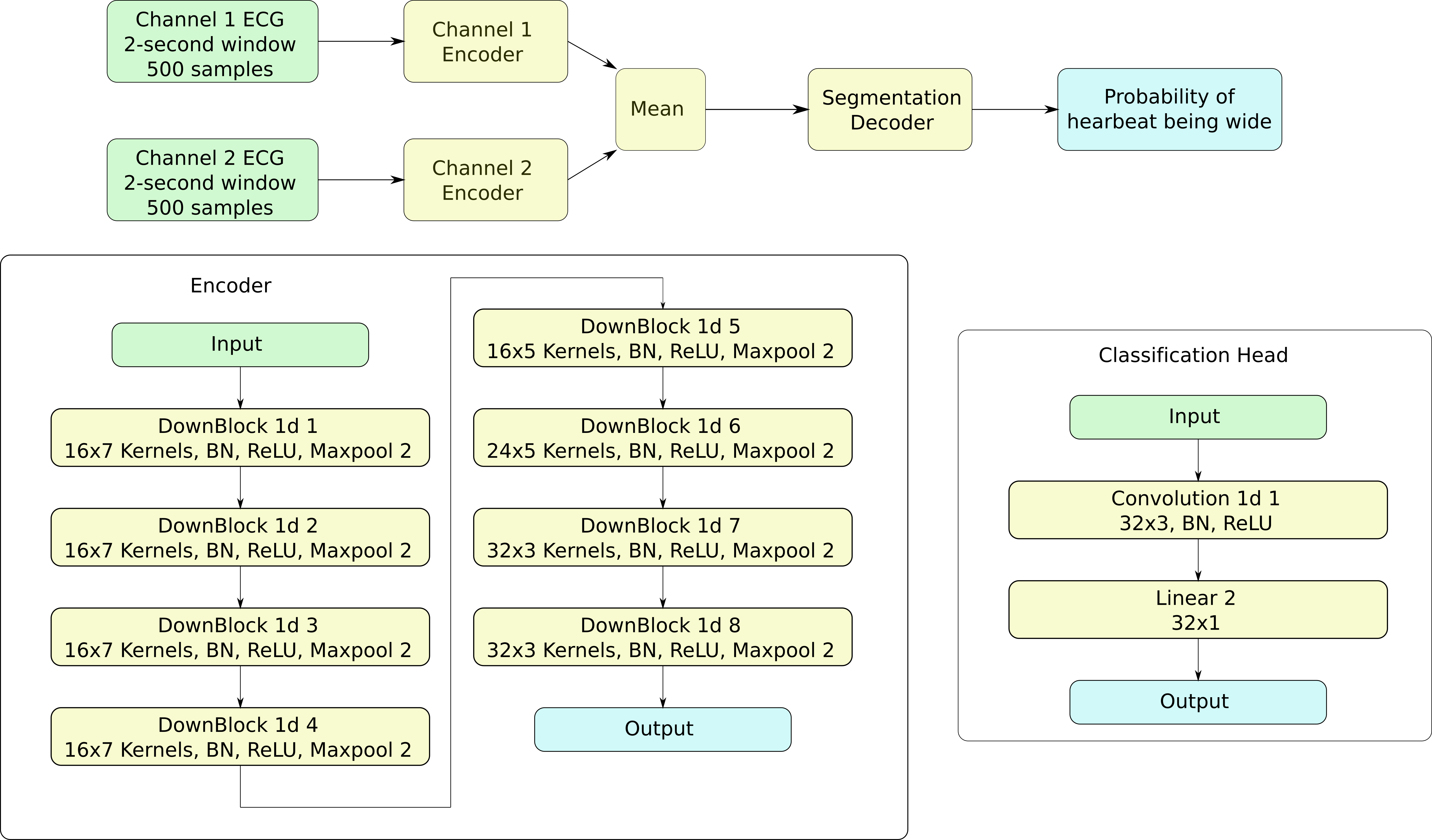, width = 15cm}}
  \caption{Neural network of the second stage. }
  \label{fig3}
\end{figure*}

\subsection{Training}

\subsubsection{Augmentation}
Augmentation is proven to be an efficient technique to improve the training process and increase the overall model performance. Our main augmentation is empirically chosen "channel-wise dropout": zeroing out one of the channels entirely with probability of 0.9 and adding strong Gaussian noise instead, thus forcing the model to learn using only the "worthy" channels. 

The rest of the augmentations we used are common in the ECG and other signal processing domains  \cite{Salamon17}: adding Gaussian noise to both channels, resampling by the factor of (0.7 - 1.3) and  channel-wise multiplication by the factor of (0.5 - 1.5).

\begin{table*}[t]
\caption{Comparison Between Different Methods for the Detection R-peaks task}\label{tab2} \centering
\begin{tabular}{|c|c|c|}
  \hline
  Method & Sensitivity (Se) & Specificity (+P) \\
  \hline
  Baseline Vendor 1 & 0.989 & 0.975 \\
  \hline
  Baseline Vendor 2 & 0.968 & 0.972 \\
 \hline
 Neural Network & { \bfseries 0.993} & 0.990  \\
  \hline
 Cardiologist & 0.991 & { \bfseries 0.992} \\
  \hline
\end{tabular}
\end{table*}

\subsubsection{Sampling and Losses}
Due to the extreme imbalance of the wide/narrow heartbeats in the dataset (almost 1:100), we applied the positive example upsampling technique during the training process. It also worth noting that various losses perform differently with different class balances. Therefore, we tested three types of losses: Binary Cross Entropy (BCE) Loss, Focal Loss, and Dice loss in both the segmentation and the classification stages. In our experiments, the best results on the validation set were achieved with the positive class ratio 3:20 in both stages, while the Dice Loss performed the best in the segmentation stage and the BCE Loss in the classification stage.

\subsubsection{Neural Networks Training}
We trained both neural networks for the 100,000 steps with the batch size of 64. We used Adam optimizer with B1 = 0.9 and B2 = 0.999, learning rate = 0.001 with exponential decay by the factor 0.97 every 1000 steps. Each batch we formed by sampling wide heartbeats with probability 0.15 and narrow with probability 0.85.

\subsubsection{GBTD Training}
The third stage classification was made by GBDT using the LightGBM library \cite{Ke17}. It was trained on the validation dataset, that was further split into 700 patients for training and 300 patients for validation in the current stage. Training was stopped after 461 iterations based on the validation set metrics. The feature set for the current stage was generated using the features obtained in the previous steps. The most important were the vectors describing each heartbeat taken from the classification network right after the channel merging step and raw outputs (logits) from the network. From these features we generated patient-wise aggregated features, such as the mean, the median and the standard deviation of each value. Further, we added features based on the heartbeat detection from the first stage. These are mean, median rate, local heartbeat rate based on nearest 100 and 10 beats and their relationship with the current heartbeat.

\begin{table*}[h]
\caption{Comparison Between Different Methods for the Classification QRS task}\label{tab2} \centering
\begin{tabular}{|c|c|c|}
  \hline
  Method & Sensitivity (Se) & Specificity (+P) \\
  \hline
  Baseline Vendor 1 & 0.687 & 0.961 \\
  \hline
  Baseline Vendor 2 & 0.318 & 0.955 \\
 \hline
 Neural Network &  0.873 & 0.997 \\
  \hline
 NN + GBDT & {\bfseries 0.917} & {\bfseries 0.999} \\
 \hline
 Cardiologist & 0.872 & {\bfseries 0.999} \\
  \hline
\end{tabular}
\end{table*}
\section{\uppercase{Results}}

The results of the conducted experiments are shown in the Table 2. As baselines for comparing our model, we chose two widely used commercial-grade software packages built by two different vendors\footnote{Due to the confidentiality agreements, we cannot reveal the names of these vendors}. Furthermore, we compare our model with the annotations made by experienced cardiologists over-reading annotations acquired with the software package produced by Vendor 2.
\begin{table*}[h]
\caption{QRS detection performance comparison on the MIT-BIH arrhythmia database}\label{tab3} \centering
\begin{tabular}{|c|c|c|}
  \hline
  Work &  Recall (\%) & Precision (\%) \\
  \hline
  Pan and Tompkins (1985) & 90.95 & 99.56 \\
  \hline
  Elgendi et al. (2009) &  87.90 & 97.60 \\
 \hline
 Chouakri et al. (2011) & 98.68 & 97.24 \\
  \hline
 Rodriguez-Jorge et al. (2014) &  96.28 & 99.71 \\
 \hline
 NN + GBDT (our work) & 98.11 & 99.91 \\
  \hline
\end{tabular}
\end{table*}
We measure performance of our models using Sensitivity (Se) and Specificity (+P) metrics. True positive for the detection task was counted if model detects a heartbeat within 150 ms of the true one. The comparison is made on the test dataset consisting of 291 recordings annotated by a committee of three cardiologists. This dataset was annotated independently from the original training dataset, which also enables us to evaluate performance of a single cardiologist.

As Table 2 demonstrates, our model outperforms the two selected baselines in the task of detecting positions of heartbeats, as well as classifying the heartbeats into narrow and wide. Moreover, the proposed model achieved the accuracy level comparable to the experienced cardiologists on these tasks, as shown in Table 2.

Furthermore, we tested our method on the MIT BIH Arrhythmia Database and compared the results of the different approaches described in the literature with the similar metrics of our model (see Table 3) \cite{Rodriguez14}. The MIT-BIH Database is a test set for evaluation of arrhythmia detection performance as well as for basic research into cardiac dynamics that has been used about 500 times worldwide since 1980 \cite{Moody01}. Due to the fact that this dataset is not suitable for the classification task, we evaluated our model only in the context of the detection task. As demonstrated in Table 3, our approach to the challenge of heartbeat detection showed stronger performance results than the selected baselines.

\section{\uppercase{Conclusions}}

In this paper we present a novel heartbeat detection and heartbeat classification (narrow or wide) method for the two-channel long-term ECGs.
We propose a channel-wise CNN architecture and combine it with the GBDT model that can employ patient-wise features.
Furthermore, we demonstrate on the set of 291 ambulatory 2-lead ECG 24-hour recordings that our method significantly outperforms two commercially available software packages widely used by the cardiologists for these tasks in the country, approaching the quality level of experienced radiologists.

As a future work, we intend to conduct prospective clinical trials for confirmation of clinical significance of this model, as well as enhancing our model for the detection and interpretation of more complex components of the heartbeat.

\bibliographystyle{apalike}
{\small
\bibliography{example}}

\end{document}